\documentclass[aps,prl,10pt,twocolumn,superscriptaddress,preprintnumbers]{revtex4}

\usepackage[dvipsnames]{xcolor}
\usepackage{physics}
\usepackage{bm}
\usepackage{braket}

\usepackage{scalerel}
\usepackage{blindtext}
\usepackage{epsfig, cancel}

\usepackage{latexsym}
\usepackage{natbib, comment}
\usepackage{mathrsfs,amsmath,amssymb,amsthm,amsfonts,tikz,graphicx,accents,hyperref,color}
\usepackage{url}
\usepackage{dcolumn}
\usepackage{multirow}
\usepackage{color}
\usepackage{cancel}
\usepackage{soul}
\usepackage[normalem]{ulem}
\usepackage{txfonts}
\usepackage{epsfig}
\usepackage{psfrag}
\usepackage{subfigure}
\hypersetup{colorlinks=true}
\usepackage{mathtools}
\usepackage{enumitem}
\usepackage{float}
\usepackage{caption,ragged2e}

\def\H0{{\text{H}\hspace*{-2.05mm}\text{H} 0\hspace*{-1.35mm}0\ }}

\usepackage[all]{xy}
\usepackage{slashed}
\usepackage{slashed,ccaption}
\usepackage{multirow}

\usepackage{caption}

\usepackage{array}
\hypersetup{ linktoc=all,
    colorlinks, linkcolor={palatinateblue},
    citecolor={red}, urlcolor={amaranth} 
}

\graphicspath{{Images/}}

\renewcommand{\d}[1]{\ensuremath{\operatorname{d}\!{#1}}}

\DeclareSymbolFont{extraup}{U}{zavm}{m}{n}
\DeclareMathSymbol{\varheart}{\mathalpha}{extraup}{86}
\DeclareMathSymbol{\vardiamond}{\mathalpha}{extraup}{87}
\makeatletter
\renewcommand*{\@fnsymbol}[1]{\ensuremath{\ifcase#1\or \clubsuit \or \vardiamond \or \varheart\or
    \spadesuit\or \mathparagraph\or \|\or **\or \dagger\dagger
    \or \ddagger\ddagger \else\@ctrerr\fi}}
\makeatother

\definecolor{rosy}{RGB}{230,235,252}
\definecolor{myframetitle}{RGB}{90,89,170}
\definecolor{myblocktitle}{RGB}{140,185,249}
\definecolor{mytitle}{RGB}{10,80,26}

\definecolor{darkgreen}{RGB}{27,130,45}
\definecolor{darkblue}{rgb}{0,0,0.3}
\definecolor{darkred}{rgb}{0.7,0,0}

\definecolor{light gray}{RGB}{220,220,220}
\definecolor{dark purple}{RGB}{108,0,217}
\definecolor{pink}{RGB}{190,20,100}
\definecolor{orang}{RGB}{193,63,0}
\definecolor{green}{RGB}{11,98,17}
\definecolor{darkpink}{RGB}{153,0,76}
\definecolor{bluegreen}{RGB}{0,102,102}
\definecolor{greenlagan}{RGB}{0,102,0}
\definecolor{redgreen}{RGB}{102,102,0}
\definecolor{Redgreen}{RGB}{153,76,0}
\definecolor{vividviolet}{rgb}{0.62, 0.0, 1.0}
\definecolor{amaranth}{rgb}{0.9, 0.17, 0.31}
\definecolor{palatinateblue}{rgb}{0.15, 0.23, 0.89}
\definecolor{brightpink}{rgb}{1.0, 0.0, 0.5}
\definecolor{cornflowerblue}{rgb}{0.39, 0.58, 0.93}
\definecolor{deepcarminepink}{rgb}{0.94, 0.19, 0.22}
\definecolor{radicalred}{rgb}{1.0, 0.21, 0.37}

\DeclareFontFamily{OT1}{rsfs}{}

\DeclareFontShape{OT1}{rsfs}{m}{n}{ <-7> rsfs5 <7-10> rsfs7 <10->rsfs10}{} 

\DeclareMathAlphabet{\mycal}{OT1}{rsfs}{m}{n}

\newcommand{\be}{\begin{equation}}
\newcommand{\ee}{\end{equation}}
\newcommand{\bec}{\begin{center}}
\newcommand{\eec}{\end{center}}

\newcommand\tcm{\textcolor{magenta}}

\newcommand{\bdry}{\mathcal{B}}
\newcommand{\cauchy}{\mathcal{C}}

\newcommand{\MB}{\mathcal{M}_\mathcal{B}}

\usepackage[most]{tcolorbox}

\tcbset{highlight math style={left=02mm,right=02mm,top=-1mm,bottom=-1mm}} 
\usepackage{empheq}

\begin{document}

\newcommand{\mytitle}{\begin{center}{\large{\textbf{{Charges in General Relativity and  Black Hole Thermodynamics}}}}
\end{center}}

\title{{\mytitle}}
\author{M.~Golshani}\email{mahdig@iasbs.ac.ir}
\affiliation{Department of Physics, Institute for Advanced Studies in Basic Sciences (IASBS),
P.O. Box 45137-66731, Zanjan, Iran}
\affiliation{School of Physics, Institute for Research in Fundamental
Sciences (IPM), P.O.Box 19395-5531, Tehran, Iran}
\author{M.~M.~Sheikh-Jabbari}\email{jabbari@theory.ipm.ac.ir}
\affiliation{School of Physics, Institute for Research in Fundamental Sciences (IPM), P.O.Box 19395-5531, Tehran, Iran}
\author{V.~Taghiloo}\email{v.taghiloo@iasbs.ac.ir}
\affiliation{School of Physics, Institute for Research in Fundamental
Sciences (IPM), P.O.Box 19395-5531, Tehran, Iran}
\affiliation{Department of Physics, Institute for Advanced Studies in Basic Sciences (IASBS),
P.O. Box 45137-66731, Zanjan, Iran}
\author{M.H.~Vahidinia}\email{ vahidinia@iasbs.ac.ir}
\affiliation{Department of Physics, Institute for Advanced Studies in Basic Sciences (IASBS),
P.O. Box 45137-66731, Zanjan, Iran}
\affiliation{School of Physics, Institute for Research in Fundamental
Sciences (IPM), P.O.Box 19395-5531, Tehran, Iran}

\begin{abstract}
 {We shed a new light on the longstanding problem of covariant charges in diffeomorphism invariant theories like General Relativity (GR) by noting the other important feature of the theory, the background independence. To this end, we develop covariant phase space formalism in which we allow  for the boundaries of spacetime to have arbitrary fluctuations. Within this formalism we show non-covariance of charges appear in inevitable integration constants which also break background independence in the expression of charges. We then apply the same formalism to black hole thermodynamics. We generalize the seminal Iyer-Wald derivation the first law of bl1ack hole thermodynamics by relaxing the need for the assumptions at a bifurcation surface and asymptotic infinity, as well as addressing questions regarding the integrability of charges. We also present a first principles derivation of the Smarr relation within our framework.} 
\end{abstract}
\maketitle


The two characteristic features of Einstein's General Relativity   {and all diffeomorphism-invariant theories of gravity} are general covariance and background independence. The former manifests itself in invariance of the theory under general coordinate transformations and the latter implies that all background geometries are solutions to the same theory; background independence is ``general covariance over the space of solutions'' of the theory. That is, GR enjoys general covariance over both the spacetime and its solution space. 

These two features obscure a covariant notion of conserved charges in GR \cite{Wald:1984rg, landau-lifshits, Grumiller:2022qhx}.  One of the objectives of this letter is to shed  {a new} light on the  {longstanding problem of} charge issue in GR. To this end, we need a suitable framework that allows us to disentangle the spacetime and solution space covariance to the extent possible. The latter is non-trivial because any solution to the theory is inevitably given by a set of scalars,  vectors, and in general tensors over the spacetime  {that transform accordingly under diffeomorphisms}. Seminal papers of  Wald et al in the early 1990s \cite{Lee:1990nz, Sudarsky:1992ty, Iyer:1994ys, Wald:1999wa} were crucial in developing covariant phase space (CPS) formalism which deals with quantities that are covariant over both solution space and spacetime and hence apt to the need. CPS formalism has been used to upgrade  Noether's theorem and charges to a more covariant version \cite{Iyer:1994ys}.   

CPS formalism yields charge variations associated with diffeomorphisms given by integrals over codimension-2 surfaces in the spacetime, as we have in the usual Gauss law for the example of electromagnetism. To be able to define the charge, the charge variation should be  ``integrable'' over the solution space. Disentangling spacetime and solution space, and hence derivatives thereon, disentangles the notion of conservation (associated with motion in time) from the integrability (associated with motion in solution space). An extension of CPS formalism, covariant phase space with fluctuating boundaries (CPSFB), has been recently proposed allowing for a full disentanglement of conservation and integrability \cite{Adami:2024gdx}. In particular, within CPSFB charges associated with diffeomorphisms are always integrable and the integrable charge is the Noether charge \cite{Adami:2024gdx}, a similar statement has also been made in \cite{Ciambelli:2021nmv, Freidel:2021dxw}.

Black holes are generic GR solutions and it is established that besides their spacetime features they behave as thermodynamical systems in the solution space. There are concise statements and derivations of the four laws of thermodynamics for stationary black holes, see e.g. \cite{Grumiller:2022qhx} and references therein. In particular, Iyer and Wald \cite{Iyer:1994ys} gave a robust derivation of the first law of black hole thermodynamics for any diffeomorphism invariant theory. The cornerstone of Iyer-Wald derivation is Wald's ``entropy as the Noether charge'' \cite{Wald:1993nt} while the other charges are not Noether charges computed at spatial asymptotic infinity of the black hole spacetime in CPS formalism. 
 {The other objective of this note is to generalize  Iyer-Wald derivation within CPSFB} and extend it to a generalized first law written in terms of Noether charges and flux. We repeat  analyses of \cite{Hajian:2015xlp} in the CPSFB framework and besides the generalized first law we provide a covariant derivation of the generalized Smarr relations \cite{Smarr:1973}. 

 {This Letter presents five notable results derived within  CPSFB formalim: (1) Integrable surface charges in gravity correspond to the covariant Noether charges, modulo  integration constants over the solution space; (2)  Noncovariance of charges are entirely coming from these integration constants; (3) We derive an ambiguity-free Gibbs-Duhem relation (Smarr formula) for black holes expressed in terms of Noether charges; (4) We explicitly demonstrate that the ambiguity-free first law of black hole thermodynamics emerges from the radial conservation of the variation of Noether charges and is valid for any cross-section of the horizon; (5) We generalize analyses of \cite{Iyer:1994ys} and derive the first law involving  Noether charges associated to any field-dependent exact symmetries (Killing vectors).}

\bec
\textbf{Review of CPSFB}\label{sec:review}
\eec

\begin{figure}
    \centering
    \includegraphics[scale=.4]{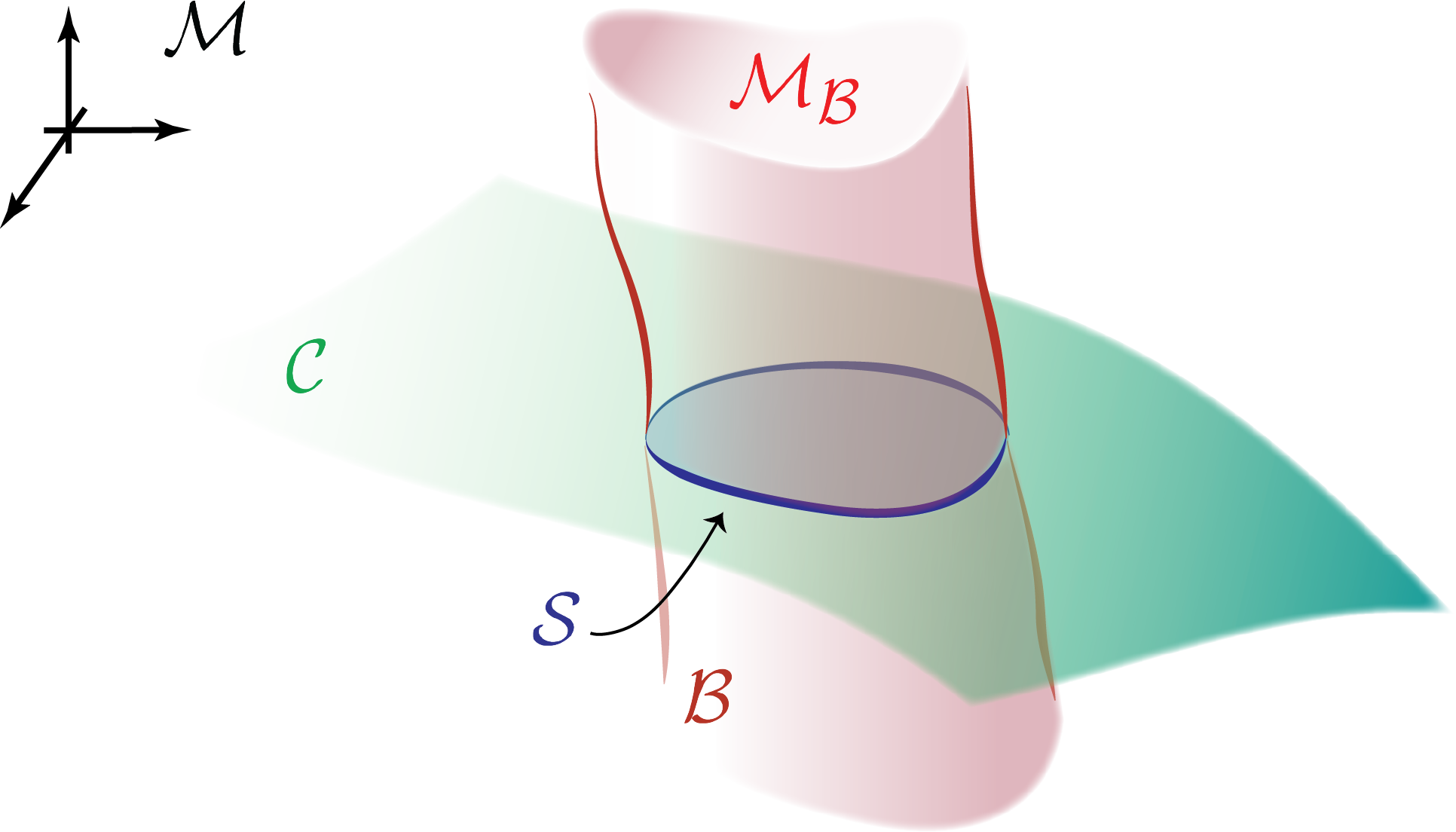}
    \caption{\justifying
         {$D$-dimensional manifold $\textcolor{red}{\MB}$ as a submanifold of a boundary-less manifold $\mathcal{M}$ which is bounded by $D-1$ dimensional timelike surface $\textcolor{BrickRed}{\bdry}$.
        $\textcolor{Green}{\cauchy}$ denotes a codimension-1 spacelike surface and intersects $\textcolor{BrickRed}{\bdry}$ over spacelike compact smooth $(D-2)$-dimensional surface $\textcolor{Blue}{{\cal S}}$.}}
    \label{fig:3dsetup}
\end{figure}
\textbf{Action and symplectic potential.}
The action for a physical system on the manifold $\MB$  {with a timelike boundary ${\cal B}(x)=0$} can be represented on the hypothetical boundary-less manifold $\mathcal{M}$  {(see FIG.~\ref{fig:3dsetup})}:
\begin{equation}\label{action-original}
	S=\int_{\MB} \d{}^{D}x\, \mathrm{L}[\Phi]=\int_{\mathcal{M}} \d{}^{D}x\, \mathrm{L}[\Phi]\,  H(\bdry) \, ,
\end{equation}
where $\mathrm{L}[\Phi]$ is Lagrangian density, $\Phi$ denotes the collection of fields and  $H(\bdry)$ is the Heaviside step function
\begin{equation}\label{step-function}
	H(\bdry):=\left\{\begin{array}{cc} 1 & \qquad \mathcal{B}(x) \geq 0\\ 0 & \qquad \mathcal{B}(x)<0 \end{array}\right. 
\end{equation}
 {Incorporating the boundary's role through $H(\mathcal{B})$ into the integral enables the proper consideration of boundary variations. }The first variation of the action, 
\begin{equation}\label{delta-S}
\hspace*{-2mm}            \delta S=\int_{\mathcal{M}} \d{}^{D}x \, \text{E}_{\Phi}\, H(\bdry)\, \delta \Phi+\int_{\bdry}\d{}x_{\mu}\, {\Theta}^{\mu}\, , \ {\Theta}^{\mu}:=\Theta^{\mu}_{\text{\tiny{LW}}}+\chi^{\mu}\, \mathrm{L}
\end{equation}
yields field equations $\text{E}_{\Phi}=0$ plus a boundary term that consists of the usual Lee-Wald symplectic potential $\Theta^{\mu}_{\text{\tiny{LW}}}$ \cite{Lee:1990nz} plus a term arising due to boundary variation  {which is described by $\chi^{\mu}$ (see \cite{Adami:2024gdx} and Supplemental Material for more details).}

The symplectic form in the fluctuating boundary setup involves the standard Lee-Wald contribution and a corner term 
\begin{equation}\label{symp-form-var-bndy}
\begin{aligned} 
\Omega[\delta \Phi, \delta \Phi; \Phi]&=\Omega_{\;\text{\tiny{LW}}}[\delta \Phi, \delta \Phi; \Phi]+\Omega_c[\delta \Phi, \delta \Phi; \Phi]\, ,\\
       \Omega_{\;\text{\tiny{LW}}}[\delta \Phi, \delta \Phi; \Phi] &:=\int_{\bdry}\d{}x_{\mu}\, \delta\Theta^\mu_{\text{\tiny{LW}}}\  H(\cauchy)\, ,  \\
       \Omega_c[\delta \Phi, \delta \Phi; \Phi] &:=
       \int_{\cal S}\d{}x_{\mu\nu}\, \big(2\, \Theta^{\mu}_{\text{\tiny{LW}}}\wedge \chi^{\nu}+\mathrm{L}\, \chi^{\mu}\wedge \chi^{\nu}\big)\, ,            
\end{aligned}
\end{equation}
where $\wedge$ denotes the wedge product over field space  { and  ${\cal C}(x)=0$  represents a spacelike (partial) Cauchy surface.  Thus the symplectic form has a $D-1$ dimensional timelike boundary $\cal{B}$ part and a $D-2$ dimensional corner ${\cal S}$ part}.

\textbf{Surface charges.}
The $(0;1)$-from \footnote{ {$(p;q)$-form  denotes a $p$-form in spacetime which is a $q$-form in the field space.}} surface charge variation associated with a generic diffeomorphism $\xi$ is  $\slashed{\delta}Q(\Phi;\xi):=\Omega[\delta_\xi \Phi, \delta \Phi; \Phi]$, where $\slashed{\delta}$ is to highlight the fact that charges are not in general integrable. Recalling the definition of the Noether current \cite{Wald:1993nt, Iyer:1994ys}
\be\label{Noether-current}
J^{\mu}_{\xi}:=\Theta^{\mu}_{\text{\tiny{LW}}}[\delta_{\xi}\Phi;\Phi]-\xi^{\mu}\text{L}=\partial_{\nu}Q_{\text{\tiny{N}}}^{\mu \nu},
\ee
for a generic $\xi$, the charge variation becomes \cite{Adami:2024gdx}
\begin{align}\label{charge-variation}
\slashed{\delta}Q &= -\int_{{\cal S}} \d{}x_{\mu\nu}\, \left(\delta Q_{\text{\tiny{N}}}^{\mu\nu}(\xi)+2J^{\mu}_{\xi}\, \chi^{\nu}\, {-\,Q_{\text{\tiny{N}}}^{\mu\nu}(\delta\xi)}\right) \nonumber\\
&=-\delta \int_{\mathcal{S}} \d{}x_{\mu \nu}Q_{\text{\tiny{N}}}^{\mu \nu}(\xi)+\int_{\cal S}\d{}x_{\mu\nu}\, Q_{\text{\tiny{N}}}^{\mu\nu}(\delta\xi)\, ,
\end{align}
where the identity 
$\delta \int_{\cal S}  \d{} x_{\mu\nu}\ X^{\mu\nu}=\int_{\cal S} \d{}^{D-2}x\, \qty(\delta X+\partial_{\mu}(X\, \chi^{\mu}))$ and $X= \frac{1}{2}\mathcal{E}_{\mu\nu}X^{\mu\nu}$  
(see \cite{Adami:2024gdx} for more details).
So,  surface charges associated with {field-independent diffeomorphism $\xi$ ($\delta\xi=0$)} within CPSFB are integrable and are given by the Noether charge, explicitly 
\begin{equation}\label{Q-QN-QR}
\hspace*{-3mm}    Q(\Phi;\xi)= - Q_{\text{\tiny{N}}}(\Phi;\xi)+Q_{_{\cal R}}(\xi), \   Q_{\text{\tiny{N}}}(\Phi; \xi):=  \int_{{\cal S}}\d{}x_{\mu\nu}\, Q_{\text{\tiny{N}}}^{\mu\nu}(\xi)
\end{equation}
where $Q_{_{\cal R}}(\xi)$ is an integration constant (over the field space), $\delta Q_{_{\cal R}}(\xi)=0$ and is linear in $\xi$. $Q_{_{\cal R}}(\xi)$ can be fixed  requiring that the reference point ${\Phi}_{_{\cal R}}$ has a vanishing charge,  $Q({\Phi}_{_{\cal R}};\xi)=0$, yielding $Q_{_{\cal R}}(\xi)=Q_{\text{\tiny{N}}}({\Phi}_{_{\cal R}};\xi)$.

\bec
\textbf{Non-covariance of charge}
\eec
 {A direct outcome of the above analysis is uncovering the origin of non-covariance in charges within any diffeomorphism-invariant theory of gravity, including GR.} To see this,  recall that $X(\Phi;\xi)$ is called covariant if \footnote{ {We restrict our analysis to field-independent diffeomorphisms ($\delta \xi=0$), though this assumption can be easily omitted if desired.}}, 
\begin{equation}\label{covariance-def}
 \hspace*{-2mm}   \delta_\eta X(\Phi;\xi):=\delta X(\Phi;\xi)\big|_{\eta} ={\cal L}_\eta X(\Phi;\xi)- X(\Phi;{\cal L}_\eta \xi)\, , \ \forall\ \xi, \eta
\end{equation}
 {where $\delta_\eta X$ and ${\cal L}_\eta X$  represent Lie derivatives of $X(\Phi;\xi)$ in solution space and spacetime respectively \cite{Wald:1999wa}. This equation implies how notion of covariance in these to spaces are entangled.}
The definitions of  $Q_{\text{\tiny{N}}}(\xi), Q_{_{\cal R}}(\xi)$ and covariance \eqref{covariance-def} yield  that $Q_{\text{\tiny{N}}}(\xi)$ is a covariant quantity, while $Q_{_{\cal R}}(\xi)$ is generically not \cite{Adami:2024gdx}:  $\delta Q_{_{\cal R}}(\xi)=0$ and hence  $\delta_\eta Q_{_{\cal R}}(\xi)=0$,  recalling covariance of Noether charge and $Q_{_{\cal R}}(\xi)=Q_{\text{\tiny{N}}}({\Phi}_{_{\cal R}};\xi)$ yields, 
\be
\begin{split}
\hspace*{-3mm} &-\delta_\eta Q_{_{\cal R}}(\xi)+{\cal L}_\eta Q_{_{\cal R}}(\xi)- Q_{_{\cal R}}({\cal L}_\eta \xi)\\ 
&=0 + {\cal L}_\eta Q_{\text{\tiny{N}}}({\Phi}_{_{\cal R}};\xi)- Q_{\text{\tiny{N}}}({\Phi}_{_{\cal R}};{\cal L}_\eta \xi)=\frac{\delta Q_{\text{\tiny{N}}}}{\delta\Phi}\big|_{_{{\Phi}_{_{\cal R}}}}\!\!\! {\cal L}_\eta{\Phi}_{_{\cal R}} \neq 0,
\end{split}
\ee
as generically ${\cal L}_\eta{\Phi}_{_{\cal R}} \neq 0$.  
This explicitly shows that \textit{the non-covariance of the charges in GR comes from integration constants in the field space $ Q_{_{\cal R}}$,  evaluated at a reference field configuration ${\Phi}_{_{\cal R}}$.} It also uncovers another important fact: since $\Phi_{_{\cal R}}$ is a solution to the theory, ${\cal L}_\eta{\Phi}_{_{\cal R}}$ is variation of $\Phi_{_{\cal R}}$ under a generic diffeomorphism $\eta$ and hence $\delta_\eta Q_{_{\cal R}}(\xi)$ is a ``soft charge'' over $\Phi_{_{\cal R}}$ in the now standard terminology of \cite{Strominger:2017zoo}. Our discussions above are a formal, more general, rigorous, and robust restatement of results in \cite{Chen:2021kug, Javadinezhad:2023mtp}.

 {While CPSFB is developed within CPS formalism, it has notable advantages over the CPS analysis usually used in the literature, see e.g. \cite{Grumiller:2022qhx}. CPSFB is ``more covariant'' as it allows for fluctuations associated with normal vectors to a given boundary and for the same reason it is also more covariant over the solution space. As a result, the only remaining source of non-covariance is the choice of reference point, which intriguingly also breaks background independence \footnote{ {We remark that, in the AdS/CFT framework, see e.g. \cite{Emparan:1999pm,Balasubramanian:1999re,Papadimitriou:2005ii}, this reference point is set by a boundary counterterm that breaks covariance, supporting our assertion that charge non-covariance originates from the choice of reference point.}}}.

\textbf{Freedoms (ambiguities).} There are two $W$ and $Y$ freedoms in the definition of symplectic potential in CPS formalism;  {$W$ is associated with adding a (total derivative) boundary term to the Lagrangian and $Y$ is coming from the fact that symplectic potential ${\Theta}^{\mu}$ is defined as on-shell variation of the action, and hence one may shift it by a total divergence \cite{Grumiller:2022qhx}. For the specific case of CPSFB \cite{Adami:2024gdx}} 
\begin{equation}\label{symp-pot-freedom}
   \hspace*{-2mm}         \tilde{\Theta}^{\mu}=\Theta^{\mu}+ \delta W^\mu[\delta\Phi;\Phi] + \chi^\mu\, \partial_\nu W^\nu[\delta\Phi;\Phi] +\partial_{\nu}Y^{\mu\nu}[\delta\Phi;\Phi] \, .
\end{equation}
One can then show that \cite{Adami:2024gdx} the freedom in defining the symplectic form and surface charge variation appears only through the combination $\bar{Y}^{\mu\nu}:=Y^{\mu\nu}-2 W^{[\mu}\ \chi^{\nu]}$, in particular,
\begin{equation}\label{charge-freedom}
    \tilde{Q}(\xi)=Q(\xi)-\int_{\cal S}\d{}x_{\mu\nu}\, (\bar{Y}^{\mu\nu}[\delta_{\xi}\Phi;\Phi]-\bar{Y}^{\mu\nu}[\delta_{\xi}\Phi_\mathcal{R};\Phi_\mathcal{R}])\, .
\end{equation}
The above shows that if $\bar{Y}$ is covariant, as defined in \eqref{covariance-def}, the non-covariance in the charge expression will still be induced only from the field-space integration constant $Q_{_{\cal R}}(\xi)$.

\bec
\textbf{CPSBF and Black Hole Thermodynamics}\label{sec:cons-theorem}
\eec
Surface charges $Q(\xi)$ \eqref{Q-QN-QR} are given by codimension-2 integrals and to discuss their conservation one should compare values of charge over two such codimension-2 surfaces ${\cal S}_1, {\cal S}_2$, which are two points on the $2d$ part of spacetime foliated by ${\cal B}, {\cal C}$, as depicted in FIG.~\ref{fig:spatial-cons}. So far $\xi$ was a generic diffeomorphism,
for the purpose of exploring laws of black hole thermodynamics, as in \cite{Iyer:1994ys}, hereafter we restrict to exact symmetries (Killing vectors) denoted by $\zeta$, i.e. $\delta_\zeta\Phi=0$ (possibly up to internal gauge transformations) \cite{Hajian:2015xlp, Grumiller:2022qhx, Barnich:2007bf}. 

We extend the analysis of \cite{Iyer:1994ys} and most of the existing literature by considering field-dependent Killing vectors $\delta\zeta\neq 0$. An explicit example of such a field-dependent Killing vector has been studied in \cite{Compere:2015knw, Sheikh-Jabbari:2016unm}. Another class of examples is the ``parametric variations'' which move us in the parameter space of black hole solutions \cite{Hajian:2015xlp, Grumiller:2022qhx}. These are the ones considered in the  original seminal work \cite{Bardeen:1973gs} and many follow-up papers.

We present four propositions, two about the conservation of Noether charges and their variation and two on their invariance under $W, Y$ choices. The proofs of these propositions are given in the supplemental material. To avoid cluttering,  among the arguments of charges and other quantities we only show those that are essential for the analyses here.

\tcm{\textbf{Proposition I (generalized Smarr relation)}.} Noether charge $Q_{\text{\tiny{N}}}(\zeta)$ associated with any exact symmetry $\zeta$ for which $\delta\zeta$ can be nonzero and $\delta_\zeta\Phi=\mathcal{L}_{\zeta}\Phi=0$, satisfies the conservation/balance equation, 
\begin{equation}\label{Noether-cons}
            \tcbset{fonttitle=\scriptsize}
            \tcboxmath[colback=white,colframe=gray]{
               Q_{\text{\tiny{N}}}(\zeta;\mathcal{S}_2)-Q_{\text{\tiny{N}}}(\zeta;\mathcal{S}_1)=-\int_{\mathcal{X}_{12}}\d{}x_{\mu}\, \zeta^{\mu}\, \mathrm{L} [\Phi]\,  ,
            }
\end{equation}
where $\mathcal{S}_1$ and $\mathcal{S}_2$ are two corners (codimension-2 surfaces) at timelike, null, or spacelike separations in the $2d$ part of spacetime, cf. FIG. \ref{fig:spatial-cons}. If timelike separated, \eqref{Noether-cons} is a temporal balance equation and if spacelike separated, it is a radial balance equation, which we will use in the next section to derive Smarr relation based on Noether charges.

\tcm{\textbf{Proposition II (generalized first law)}.} For any field or parameter dependent exact symmetry $\zeta$, (0;1)-form  $K(\zeta;\mathcal{S})$
\begin{equation}\label{K-delta-nonzero}
             K(\zeta;\mathcal{S}):=\delta\!\!\! \int_{\mathcal{S}}\d{}x_{\mu\nu}\, Q^{\mu\nu}_{\text{\tiny{N}}}(\zeta)-\int_{\mathcal{S}}\d{}x_{\mu\nu}\, Q^{\mu\nu}_{\text{\tiny{N}}}(\delta\zeta)+2\!\! \int_{\mathcal{S}}\d{}x_{\mu\nu}\, \zeta^{\mu}\, \Theta^{\nu}
\end{equation}
is independent of the position of the corner $\mathcal{S}$. In other words, for any two codimension-2 surfaces ${\cal S}_1, {\cal S}_2$ (cf. FIG.~\ref{fig:spatial-cons}),
\begin{equation}\label{temp-rad-cons}
            \tcbset{fonttitle=\scriptsize}
            \tcboxmath[colback=white,colframe=gray]{
             K(\zeta;\mathcal{S}_1)=K(\zeta;\mathcal{S}_2)\, .
            }
\end{equation}
If ${\cal S}_1, {\cal S}_2$ are timelike separated, $K(\zeta;\mathcal{S})$ is a conserved quantity in time, and if spacelike separated, it is radially conserved. We will derive the first law of black hole thermodynamics as a radial conservation of the above quantity. We note that the first two terms in \eqref{K-delta-nonzero} are in terms of Noether charges while the last term may be viewed as a Noether flux, cf. \eqref{Noether-current}.

\tcm{\textbf{Proposition III}.} The generalized Smarr relation \eqref{Noether-cons} is independent of $W$ and $Y$ freedoms/ambiguities. 

\begin{figure}
    \centering
    \begin{tikzpicture}
\fill[black!5] (0,0.5) rectangle (5,3);
\node at (0.6,2.4) {$\bdry \cauchy$};
\draw[->] (0.2,2.2) -- (0.45,2.7);
\draw[->] (0.2,2.2) -- (0.9,2.2);
\draw[thick,darkred] (0.7,0.8) to [out=-10,in=180] (2.5,1.5);
\draw[thick,darkred] (2.5,1.5) to [out=0,in=175] (4.4,2.5);
\node[darkred] at (2.4,1.9) {$\mathcal{X}_{12}$};
\fill[Blue] (0.7,0.8) circle (2pt) node[above] {$\mathcal{S}_1$};
\fill[Blue] (4.4,2.5) circle (2pt) node[above] {$\mathcal{S}_2$};
\end{tikzpicture}
    \caption{\justifying $\textcolor{Blue}{\mathcal{S}_1}, \textcolor{Blue}{\mathcal{S}_2}$ are two generic compact codimension-2 surfaces on spacetime $\MB$, while two points on the (1+1)-dimensional ${\cal B}{\cal C}$ plane that is the time-radius plane usually used to draw Penrose diagrams. $\textcolor{Blue}{\mathcal{S}_1}, \textcolor{Blue}{\mathcal{S}_2}$ may be timelike (null or spacelike) separated and \textcolor{darkred}{${\cal X}_{12}$} is a timelike (null or spacelike) codimension-1 surface connecting $\textcolor{Blue}{\mathcal{S}_1}, \textcolor{Blue}{\mathcal{S}_2}$. } 
    \label{fig:spatial-cons}
\end{figure}

\tcm{\textbf{Proposition IV}.}  For exact symmetries  $K(\zeta;\cal S)$ and hence the generalized first law  \eqref{temp-rad-cons} are  invariant under both $W$ and $Y$ freedoms.

\bec
\textbf{First Law from Noether Charges}\label{sec:first-law}
\eec
Consider a stationary, axisymmetric spacetime with a bifurcate Killing horizon $\text{H}$ and denote the corresponding Killing vectors respectively by $\zeta_t$, $\zeta_\phi$ and 
\begin{equation}\label{zeta-H}
\zeta_{\text{\tiny{H}}}=\frac{2\pi}{\kappa_{\text{\tiny{H}}}}(\zeta_t{+}\Omega_{\text{\tiny{H}}}\, \zeta_\phi)\, ,\quad  \zeta_{\text{\tiny{H}}}^2\big|_{{\text{\tiny{H}}}}=0\, ,    \quad  \zeta_{\text{\tiny{H}}}\big|_{\mathcal{S}_{_{b}}}=0\, ,    
\end{equation} 
where $\kappa_{\text{\tiny{H}}}$ and $\Omega_{\text{\tiny{H}}}$ are respectively horizon surface gravity and angular velocity,  $\zeta_{\text{\tiny{H}}}$ is the horizon generating Killing vector and let $\mathcal{S}_{_{b}}$ be the bifurcation point.

The rest of the derivation is similar to that of \cite{Iyer:1994ys} and we briefly review here. Consider \eqref{temp-rad-cons} which is true for any ${\cal S}_1, {\cal S}_2$ and write it for $\mathcal{S}_{_{b}}$ and ${\cal S}_{i^0}$, asymptotic spacelike infinity:
\begin{equation}\label{r-cons-1}
K(\zeta_{\text{\tiny{H}}};\mathcal{S}_{_{b}})=K(\zeta_{\text{\tiny{H}}};{\cal S}_{{i^0}})\, ,
\end{equation}
and compute the LHS and RHS of the above:
\begin{equation}
     K(\zeta_{\text{\tiny{H}}};\mathcal{S}_{_{b}})
     =\delta S_{\text{\tiny{H}}}\, ,
\end{equation}
$\zeta_{\text{\tiny{H}}}^{\mu}\, \Theta^{\nu}$ term is vanishing using the fact that $\zeta^{\mu}_{\text{\tiny{H}}}\big|_{{\mathcal{S}_{_{b}}}}=0$, $S_{\text{\tiny{H}}}$ in the last equality  is the Wald's entropy \cite{Wald:1993nt}, and 
\begin{equation}
    \begin{split}
        &K(\zeta_{\text{\tiny{H}}};\mathcal{S}_{i^0})
        =\delta \left[\frac{1}{T_{\text{\tiny{H}}}}Q_{\text{\tiny{N}}}(\zeta_{t};\mathcal{S}_{i^0})\right]+\delta \left[\frac{\Omega_{\text{\tiny{H}}}}{T_{\text{\tiny{H}}}}Q_{\text{\tiny{N}}}(\zeta_{\phi};\mathcal{S}_{i^0})\right]\\
        &-\delta \big(\frac{1}{T_{\text{\tiny{H}}}}\big)Q_{\text{\tiny{N}}}(\zeta_{t};\mathcal{S}_{i^0})-\delta\left(\frac{\Omega_{\text{\tiny{H}}}}{T_{\text{\tiny{H}}}}\right)Q_{\text{\tiny{N}}}(\zeta_{\phi};\mathcal{S}_{i^0})
        +\frac{2}{T_{\text{\tiny{H}}}}\int_{\mathcal{S}_{i^0}}\d{}x_{\mu\nu}\, \zeta_{t}^{\mu}\, \Theta^{\nu}\\
        &=\frac{1}{T_{\text{\tiny{H}}}}\left[\delta Q_{\text{\tiny{N}}}(\zeta_{t};\mathcal{S}_{i^0})+2\int_{\mathcal{S}_{i^0}}\d{}x_{\mu\nu}\, \zeta_{t}^{\mu}\, \Theta^{\nu}\right]
        +\frac{\Omega_{\text{\tiny{H}}}}{T_{\text{\tiny{H}}}} \, \delta Q_{\text{\tiny{N}}}(\zeta_{\phi};\mathcal{S}_{i^0})\, ,\nonumber
    \end{split}
\end{equation}
where we used that
$T_{\text{\tiny{H}}}:=\kappa_{\text{\tiny{H}}}/2\pi$ is the Hawking temperature \cite{Hawking:1974rv} and $\d{}x_{\mu\nu}$ has non-zero components only in ${\cal B}{\cal C}$ plane, cf. \eqref{bi-normal}, \eqref{notation-int}, $\d{}x_{\mu\nu} \zeta_\phi^{\mu}=0$.
Note that unlike \cite{Iyer:1994ys} in our derivation  we include parametric variations, i.e. $\delta T_{\text{\tiny{H}}}, \delta \Omega_{\text{\tiny{H}}}$ are nonzero. 
Following \cite{Iyer:1994ys}, we have 
\begin{equation}
    \begin{split}
        &\delta Q_{\text{\tiny{N}}}(\zeta_{t};\mathcal{S}_{i^0})+2\int_{\mathcal{S}_{i^0}}\d{}x_{\mu\nu}\, \zeta_{t}^{\mu}\, \Theta^{\nu}=\delta M\, ,\\
        &\delta Q_{\text{\tiny{N}}}(\zeta_\phi;\mathcal{S}_{i^0})={-}\delta J\, ,
    \end{split}
\end{equation}
where $M$ and $J$ are the ADM mass and angular momentum respectively.
The first law of black hole thermodynamics is 
\begin{equation}\label{first-law}
    \delta M=T_{\text{\tiny{H}}}\, \delta S_{\text{\tiny{H}}}+\Omega_{\text{\tiny{H}}}\, \delta J\, .
\end{equation}
Since our derivation of the first law is based on $K(\zeta; {\cal S})$, even though it uses Noether charges, it is not affected by choice for $W, Y$ freedoms \footnote{ {We remark that, unlike Euclidean methods which rely on selecting a specific boundary Lagrangian (corresponding to a choice of $W$ freedom in our notation) e.g. \cite{Emparan:1999pm}, our derivation of the first law holds irrespective of the boundary Lagrangian chosen.}}. We close this part by the comment that the generalized first law \eqref{temp-rad-cons} may be written for any ${\cal S}_1, {\cal S}_2$ (besides $\mathcal{S}_b, \mathcal{S}_{i^0}$), all of them leading to a first law  {(see also \cite{Wald:1984rg,Jacobson:1993vj, Mishra:2017sqs})}. 

\bec
\textbf{Smarr Formula from Noether Charges}\label{sec:Smarr-law}
\eec
Smarr formula for black holes may be viewed as the Gibbs-Duhem equation in thermodynamics. While it has been extensively discussed in the literature, a first-principle ambiguity-free derivation of it is still lacking (for earlier efforts see e.g. \cite{Barnich:2004uw, Liberati:2015xcp}). In CPSFB all the charges are integrable and it hence provides the framework for a derivation of the Smarr relation. Consider \eqref{Noether-cons} for  $\zeta=\zeta_{\text{\tiny{H}}}$ and ${\cal S}_1=\mathcal{S}_{_{b}}\, , \,{\cal S}_2={\cal S}_{i^0}$\,,
\begin{equation}\label{smarr-with-QN}
    Q_{\text{\tiny{N}}}(\zeta_{\text{\tiny{H}}};\mathcal{S}_{i^{0}})-Q_{\text{\tiny{N}}}(\zeta_{\text{\tiny{H}}};\mathcal{S}_{_{b}})=-\int_{\cauchy_{12}}\d{}x_{\mu}\, \zeta_{\text{\tiny{H}}}^{\mu}\, \mathrm{L}[\Phi]\, ,
\end{equation}
where $\cauchy_{12}$ is a (partial) Cauchy surface connecting $\mathcal{S}_{_{b}}$ and ${\cal S}_{i^0}$.
 Straightforward computation yields,
\begin{equation}\begin{split}
    Q_{\text{\tiny{N}}}(\zeta_{\text{\tiny{H}}};\mathcal{S}_{_{b}})&=S_{\text{\tiny{H}}}\,,\\
     T_{\text{\tiny{H}}}   Q_{\text{\tiny{N}}}(\zeta_{\text{\tiny{H}}};\mathcal{S}_{i^0}) & = Q_{\text{\tiny{N}}}(\zeta_t;\mathcal{S}_{i^0})\,+\,\Omega_{\text{\tiny{H}}}\, Q_{\text{\tiny{N}}}(\zeta_\phi;\mathcal{S}_{i^{0}})\,  
     \\
    & = M_{\text{\tiny{N}}}-\Omega_{\text{\tiny{H}}}\, J\,,
\end{split}
\end{equation}
where $M_{\text{\tiny{N}}}=Q_{\text{\tiny{N}}}(\zeta_t;\mathcal{S}_{i^{0}})$ is the Noether mass, and finally, 
\begin{equation}\label{Smarr-1}
            \tcbset{fonttitle=\scriptsize}
            \tcboxmath[colback=white,colframe=gray]{
             -M_{\text{\tiny{N}}}+\Omega_{\text{\tiny{H}}}\, J+T_{\text{\tiny{H}}}\, S_{\text{\tiny{H}}}= T_{\text{\tiny{H}}}\int_{\cauchy_{12}}\d{}x_{\mu}\, \zeta_{\text{\tiny{H}}}^{\mu}\, \mathrm{L}[\Phi]\, . 
            }
\end{equation}
Some comments are in order: (1) Noether mass  $M_{\text{\tiny{N}}}$ and ADM mass $M$ that appear in the first law \eqref{first-law} are different;
(2) The Smarr relation \eqref{Smarr-1} and the first law \eqref{temp-rad-cons} only involve covariant quantities (and not non-covariant $Q_{_{\mathcal{R}}}$'s); (3) Unlike the first law, the RHS of \eqref{Smarr-1} is not universal (in charges) and depends on the theory \footnote{One may ask if the RHS of  \eqref{Smarr-1} can be written as a surface integral, as in the LHS. See \cite{Ortin:2021ade, Mitsios:2021zrn} for a discussion.}; (4)  Proposition III guarantees that the Smarr formula is free of $W$ and $Y$ freedoms/ambiguities; (5)  {The Smarr formula can be written in terms of ADM mass $M=M_N+2\int \d{}{x}_{\mu \nu}\, \zeta^{\mu}_{t}B^{\nu}$ as follows}
\begin{equation}\label{Smarr-3}
            \tcbset{fonttitle=\scriptsize}
            \tcboxmath[colback=white,colframe=gray]{
           \hspace{-.35 cm} -M+\Omega_{\text{\tiny{H}}}\, J+T_{\text{\tiny{H}}}\, S_{\text{\tiny{H}}}= T_{\text{\tiny{H}}}\int_{\cauchy_{12}}\d{}x_{\mu}\, \zeta_{\text{\tiny{H}}}^{\mu}\, (\mathrm{L}[\Phi]- \partial_\nu B^\nu)\,. \hspace{-.35 cm} 
            }
\end{equation}
 {where $B$-term is defined as follows \cite{Iyer:1994ys,Iyer:1995kg}
\begin{equation}\label{Theta-B}
    \int_{\mathcal{S}_{i^0}}\d{}x_{\mu\nu}\, \zeta_t^{\mu}\, \Theta^{\nu}=\delta\, \int_{\mathcal{S}_{i^0}}\d{}x_{\mu\nu}\, \zeta_t^{\mu}\, B^{\nu}\, .
\end{equation}
Proof of \eqref{Smarr-3} is given in supplementary material. This equation clarifies the role of the $B$-term in the seminal work of Iyer-Wald \cite{Iyer:1994ys}: It appears as the boundary Lagrangian.
RHS of \eqref{Smarr-2} may be viewed as a (Euclidean) on-shell action $S_{\text{E}}$, and hence it may be rewritten as 
$M-\Omega_{\text{\tiny{H}}}J-T_{\text{\tiny{H}}}\, S_{\text{\tiny{H}}}=T_{\text{\tiny{H}}}\, S_{\text{E}}$, where ${T}_{\text{\tiny{H}}}\,S_{\text{E}}$ is the free energy \cite{Gibbons:1976ue} (see also \cite{Iyer:1995kg}). }
\paragraph{\textbf{Example.}}\label{sec:example}

As an illustrative example of  our rigorous derivations, we discuss 4 dimensional ($4d$) Kerr-(A)dS black hole which is a solution to Einstein's gravity with the Lagrangian and field equations 
\begin{equation}
     \mathrm{L}= \frac{\sqrt{-g}}{16 \pi G}(R-2\Lambda)\, , \qquad R_{\mu\nu}-\frac{1}{2}R\,g_{\mu\nu}+\Lambda\, g_{\mu\nu}=0\, ,
\end{equation}
where on-shell $ \mathrm{L}= \sqrt{-g}\frac{\Lambda}{8 \pi G}$.
Symplectic potential and the Noether charge are respectively given by
\begin{equation}
\Theta^{\mu}_{\text{\tiny{LW}}}=\frac{\sqrt{-g}}{8 \pi G}\nabla^{[\alpha}(g^{\mu] \beta}\delta g_{\alpha\beta})\, , \quad Q_{\text{\tiny{N}}}^{\mu\nu}(\xi)=-\frac{\sqrt{-g}}{8 \pi G}\nabla^{[\mu}\xi^{\nu]}\, .
\end{equation}
The $4d$ Kerr-(A)dS black hole metric is given by
\begin{equation}\label{Kerr-AdS metric}
    \begin{split}
       &\hspace{-4mm}\d{}s^{2} = -\Delta_\theta\left(\frac{1-\Lambda r^2/3}{\Xi}-\Delta_\theta\, f\right)\d{}t^2-2\,\Delta_\theta\, f\,a\,\sin ^2 \theta\,\d{}t \d{}\phi\\
      &\hspace{-1mm}+\frac{\rho ^2}{\Delta_r}\d{}r^2+\frac{\rho ^2}{\Delta_\theta} \d{}\theta ^2 +\left( \frac{r^2+a^2}{\Xi}+fa^2\sin ^2\theta \right)\sin ^2\theta\,\d{}\phi ^2\,,
    \end{split}
\end{equation}
with  
\begin{equation}
    \begin{split}
        &\hspace{-2mm}\Delta_r=\left(r^2+a^2\right)\left(1-\frac{\Lambda r^2}{3}\right)-2{G} m\, r\, ,\hspace{2mm}  \Xi=1+\frac{\Lambda a^2}{3}\,,\\
        &\hspace{-2mm}\rho^2=r^2+a^2\, \cos^2\theta\, , \ \Delta_{\theta}=1+\frac{\Lambda a^2}{3}\, \cos^2\theta\, , \ f=\frac{2{G} m r}{\rho^2 \Xi^2}\, ,
    \end{split}
\end{equation}
where $m$ and $a$ are related to the ADM mass {$K(\zeta_{t},\mathcal{S}_{i^0})=\delta M$} and angular momentum {$Q_{\text{\tiny{N}}}(\zeta_{\phi},\mathcal{S}_{i^0})=-J$} as
\begin{equation}
    M=\frac{ m}{\Xi^2}\, , \qquad J=\frac{ m\, a}{\Xi^2}\, .
\end{equation}

The Hawking temperature, horizon angular velocity, and the Bekenstein-Hawking (Wald) entropy are given by
\begin{equation}
    \begin{split}
        & T_{\text{\tiny{H}}}=\frac{r_{\text{\tiny{H}}}(1-\Lambda \, a^{2}/3-\Lambda\, r_{\text{\tiny{H}}}^2-a^2\, r_{\text{\tiny{H}}}^{-2})}{4\pi \,(r_{\text{\tiny{H}}}^2+a^2)}\, ,\\
        & \Omega_{\text{\tiny{H}}}=\frac{a\,(1 -\Lambda \, r_{\text{\tiny{H}}}^2/3)}{(a^2+r_{\text{\tiny{H}}}^2)}\,, \quad S_{\text{\tiny{H}}}=\frac{\pi\, (a^2+r_{\text{\tiny{H}}}^2)}{{G}\ \Xi}\, ,
    \end{split}
\end{equation}
where $r_{\text{\tiny{H}}}$ is the horizon radius with $\Delta_r(r_{\text{\tiny{H}}})=0$. 
The analysis of the first law for this case may be found in \cite{Grumiller:2022qhx} and here we consider the Smarr relation. We note that
\begin{equation}\label{MN-Kerr-AdS-4d}
    M_{\text{\tiny{N}}}=M\left(1-\frac{\Xi}{2}\right)-\frac{\Lambda r_{\infty}(a^2+r_{\infty}^2)}{6 \,G\, \Xi}\, ,
\end{equation}
and
\begin{equation}\label{int-L-Kerr-AdS-4d}
    T_{\text{\tiny{H}}}\int_{\cauchy_{12}}\d{}x_{\mu}\, \zeta_{\text{\tiny{H}}}^{\mu}\, \mathrm{L}[\Phi]=-\frac{\Lambda r_{\text{\tiny{H}}}(a^2+r_{\text{\tiny{H}}}^2)}{6\,{G}\, \Xi}+\frac{\Lambda r_{\infty}(a^2+r_{\infty}^2)}{6\,{G}\, \Xi}\, .
\end{equation}
While  both sides of the Smarr relation \eqref{MN-Kerr-AdS-4d} and \eqref{int-L-Kerr-AdS-4d} are divergent, they cancel out and the Smarr relation is finite:
\begin{equation}
    M\left(1-\frac{\Xi}{2}\right)=\Omega_{\text{\tiny{H}}}\, J+\left(T_{\text{\tiny{H}}}+\frac{\Lambda r_{\text{\tiny{H}}} }{6\pi }\right)\, S_{\text{\tiny{H}}} \, .
\end{equation}
In the $\Lambda\to 0$ flat limit, we get the Smarr relation for Kerr black hole ${M}=2(\Omega_{\text{\tiny{H}}}\, J+T_{\text{\tiny{H}}}\, S_{\text{\tiny{H}}})$ \cite{Smarr:1973}. In the supplemental material we have analyzed  (A)dS-Myers-Perry black hole in D-dimensions as another example.
\bec
\textbf{Discussion}\label{sec:conc}
\eec
We uncovered some of the physical implications of CPSFB formalism developed in \cite{Adami:2024gdx} that all surface charges are integrable equal to the Noether charge in any diffeomorphism invariant and background independent theory, GR or beyond. We discussed how CPSFB helps to separate the two interconnected notions of charge integrability and conservation which in turn paves the way to clarify the root of non-covariance and/or background non-independence of the charge as an integration constant over the solution space.  

We proposed generalized first law \eqref{temp-rad-cons} and Smarr relations \eqref{Noether-cons} for generic stationary black holes written in terms of Noether charges and flux;  \eqref{Noether-cons} and \eqref{temp-rad-cons} are free of $W, Y$ freedoms/ambiguities. The key in our derivation of the generalized first law is the $(0;1)$-form $K(\zeta; {\cal S})$ \eqref{K-delta-nonzero} in which $\zeta$ can be a field or parameter dependent Killing vector and  ${\cal S}$ is an arbitrary codimension-2 surface. One can show that  $K$ is the unique $(0;1)$-form that is linear in $\zeta$ and is invariant under $W, Y$ choices. In our analysis we chose spacelike separated ${\cal S}_1, {\cal S}_2$ while in \eqref{Noether-cons} and \eqref{temp-rad-cons} they can also be lightlike or timelike separated, e.g. they can be any two points on the Killing horizon and asymptotic null infinity or any two points on AdS boundary. It is instructive to explore ${\cal S}_1, {\cal S}_2$ null or timelike separated cases more thoroughly.  

Our main results are based on the integrability of charges in the CPSFB framework which is applicable to non-stationary dynamical black holes. This formalism allows us to rederive and extend the analysis in recent papers \cite{Hollands:2024vbe}. We hope to explore this problem in our upcoming work.

\textbf{Acknowledgement.} We thank Hamed Adami for long term collaboration on related topics, especially on \cite{Adami:2024gdx} which this work is based on. We would also like to thank Stefano Liberati and  members
of IPM HEP-TH weekly group meeting. MG acknowledges partial support of IPM funds.  MMShJ is supported in part by Iran National Science Foundation (INSF) Grant No. 4026712 and in by part the ICTP Senior Associates Programme (2023-2028) and by ICTP HECAP section. 
\appendix 
\bec
\textbf{Supplemental Material}
\eec
Here,  {we introduce the mathematical setup,} present proofs of the 4 propositions, an alternative form for the Smarr relation, extend the proof of the first law to include parametric variations and discuss the Smarr relation for $D$ dimensional (A)dS-Myers-Perry black holes. 
\bec
\textbf{Mathematical setup}
\eec
 Consider a $D$-dimensional spacetime spanned by coordinates $x^\mu$ and metric $g_{\mu\nu}$. Let ${\cal B}(x)=0$ and ${\cal C}(x)=0$ respectively denote a codimension-1 timelike boundary ${\cal B}$ and a (partial) Cauchy surface ${\cal C}$ in it,  $\MB$  be a part of spacetime in  ${\cal B}(x)\geq 0$, and ${\cal C}(x)>0$ ($<0$) be the part of spacetime in future (past) of ${\cal C}$.  The timelike boundary ${\cal B}$, and (partial) Cauchy surface ${\cal C}$ intersect on a spacelike codimension-2 compact and smooth surface $\mathcal{S}$, see FIG.~\ref{fig:3dsetup}. One may view ${\cal C}$ as a constant time and ${\cal B}$ as a constant radius slice and  $\mathcal{S}$ as a (topologically) $S^{D-2}$ geometry.

All functions in our calculations are assumed to be smooth over ${\cal S}$ and we drop integrals of total divergences on ${\cal S}$. Vectors $\partial_\mu{\cal B}, \partial_\mu{\cal C}$, with norms $N_{\cal B}, N_{\cal C}$ are respectively spacelike and timelike and are not necessarily orthogonal to each other; one may define ${\cal A}:={N}_{\cal B}^2\,\partial_\mu{\cal B}\,\partial^\mu{\cal C}$. Induced metrics on the boundary $\bdry$ and corner $\mathcal{S}$ are 
respectively defined as
\begin{subequations}
    \begin{align}
       \hspace*{-2mm}  \gamma_{\mu\nu}&:=g_{\mu\nu}- b_\mu b_\nu\, , \qquad b_\mu  = -\, N_{\bdry} \, \partial_\mu \bdry \, , \label{induced-Sigma}\\
       \hspace*{-2mm} q_{\mu\nu}& :=\gamma_{\mu\nu}+ n_{\mu}\, n_{\nu}\, , \qquad  n_\mu =- N  (\partial_\mu \cauchy - \mathcal{A} \, \partial_\mu \bdry )\, , \label{induced-Sigmac}
    \end{align}
\end{subequations}
where $b_\mu b^\mu=1$ and $n_\mu n^\mu=-1$.
The three components of the 2-dimensional ($2d$) metric transverse to ${\cal S}$ are parametrized by ${\cal A}(x), {\cal B}(x), {\cal C}(x)$. The square root of the determinant of the above metrics are,  
\begin{equation}
\sqrt{-g}= N_\bdry \, \sqrt{- \gamma}={N}\,  N_\bdry \, \sqrt{q}\, .  
\end{equation}
${\cal E}_{\mu\nu}$, the  bi-normal to  the codimension-2 surface $\mathcal{S}$, is 
\begin{equation}\label{bi-normal}
\begin{split}
        {\cal E}_{\mu\nu}:
        =2 \partial_{[\mu}\cauchy\, \partial_{\nu]}\bdry\, ,\quad \mathcal{E}_{\mu \nu} \mathcal{E}^{\mu \nu} = -2 N^{-2} N_\bdry^{-2},
\end{split}
\end{equation}
where we use the antisymmetrization convention $X_{[\mu}\, Y_{\nu]}=\frac12(X_{\mu}Y_{\nu}-X_{\nu}Y_{\mu})$. Given the above we can define measure of integrals over $\bdry$, $\cauchy$ and $\mathcal{S}$:
\begin{equation} \label{notation-int}
\begin{split}
   \hspace{-3mm}\int_{\mathcal{X}}\d{}x_{\mu}:=-\int_{\cal X}\, \d{}^{D-1}x\, \partial_{\mu}{\mathcal{X}}\,, \hspace{1.5mm}
\int_{\cal S}\d{}x_{\mu\nu}&:=\frac{1}{2}\int_{\cal S}\, \d{}^{D-2}x\, {\cal E}_{\mu\nu}\,,
\end{split}
\end{equation}
where $\mathcal{X}$ can be $ \bdry$ or $\cauchy$.

\textbf{$(p;q)$-forms.} Besides the geometric quantities above we deal with fields on spacetime, generically denoted by $\Phi$. The variations $\delta\Phi$ denote 1-forms on the field space. We use the notation $(p;q)$-form to denote a $p$-form in spacetime which is a $q$-form in the field space. Hence ${\cal E}_{\mu \nu}$ is a (2;0)-form, $\delta{\cal A}, \delta{\cal B},\delta{\cal C}$ are (0;1)-forms.  A key quantity appearing in the CPSFB formalism is the $(1;1)$-form $\chi^{*}_{\mu}$,
\begin{equation}\label{chi-chi*-def}
\chi^*_{\mu}:=\partial_\mu \bdry\, \delta \cauchy-\partial_\mu \cauchy\, \delta \bdry\, ,
\end{equation}
one may also define the vector $\chi^{\mu}:=N^2 N_\bdry^2 \,{\cal E}^{\mu\nu}\chi^{*}_{\nu}$, which is a 1-form over the solution space. 


\bec
\textbf{\tcm{Proof of Proposition I.}} 
\eec
We start with evaluating the difference of the Noether charge at two corners. We assume these corners are connected by a codimension-1 hypersurface $\mathcal{X}_{12}$ (see FIG.~\ref{fig:3}). We take $\mathcal{X}_{12}$ to be spacelike/timelike/lightlike hypersurface if two corners are spacelike/timelike/lightlike separated. To this end, we have
\begin{equation}\label{proof-I}
   \begin{split}
        Q_{\text{\tiny{N}}}(\mathcal{S}_2)-Q_{\text{\tiny{N}}}(\mathcal{S}_1) & =\int_{\mathcal{S}_2}\d{}x_{\mu\nu}\, Q^{\mu\nu}_{\text{\tiny{N}}}(\xi)-\int_{\mathcal{S}_1}\d{}x_{\mu\nu}\, Q^{\mu\nu}_{\text{\tiny{N}}}(\xi)\\
        & =\int_{\mathcal{X}_{12}}\d{}x_{\mu}\, \partial_{\nu}Q_{\text{\tiny{N}}}^{\mu\nu}(\xi)= \int_{\mathcal{X}_{12}}\d{}x_{\mu}\, J_{\xi}^{\mu}\\
       & =\int_{\mathcal{X}_{12}}\d{}x_{\mu}\, (\Theta_{\text{\tiny{LW}}}^{\mu}[\delta_{\xi}\Phi;\Phi]-\xi^{\mu}\, \mathrm{L})\, .
   \end{split}
\end{equation}
The above is  for generic diffeomorphisms $\xi$. For an  exact symmetry $\zeta$, $\delta_\zeta\Phi=0$, $\Theta_{\text{\tiny{LW}}}^{\mu}[\delta_{\zeta}\Phi;\Phi]=0$, hence we arrive at \eqref{Noether-cons}.

\begin{figure}[h]
    \centering
    \begin{tikzpicture}[scale=0.65]
\fill[black!5] (-0.7,-0.2) rectangle (7.7,4);
\node at (-0.03,3.43) {$\mathcal{B} \mathcal{C}$};
\draw[->] (-0.55,3.15) -- (-0.25,3.8);
\draw[->] (-0.55,3.15) -- (0.3,3.15);
\draw[thick,darkred] (-0.4,1.25) to [out=-5,in=175] (1.3,1.1);
\draw[thick,darkpink] (1.3,1.1) to [out=-5,in=185] (6,2.2);
\draw[thick,darkred] (6,2.2) to [out=5,in=170] (7.4,2.0);
\node[darkpink] at (3.45,1.85) {$\mathcal{X}_{12}$};
\node[darkred] at (-0.1,1.55) {$\mathcal{X}$};
\draw[thick,darkgreen] (7.1,0.3) to [out=125,in=295] (6,2.2);
\draw[thick,darkgreen] (6,2.2) to [out=115,in=300] (5.4,3.5);
\node[darkgreen] at (6.7,0.4) {$\mathcal{Y}_{2}$};
\draw[thick,darkgreen] (1.7,0.1) to [out=115,in=290] (1.3,1.1);
\draw[thick,darkgreen] (1.3,1.1) to [out=110,in=290] (0.4,3.2);
\node[darkgreen] at (1.3,0.3) {$\mathcal{Y}_{1}$};
\fill[Blue] (1.3,1.1) circle (2pt); \node[Blue] at (1.55,1.35) {$\mathcal{S}_1$};
\fill[Blue] (6,2.2) circle (2pt);
\node[Blue] at (6.3,2.45) {$\mathcal{S}_2$};
\end{tikzpicture}
    \caption{\justifying
    The $2d$ ${\cal B}{\cal C}$ plane with 2 codimension-1 boundaries $\textcolor{darkgreen}{{\cal Y}_i}, i=1,2$ and the codimension-1 line $\color{darkred}{\cal X}$ we choose $\textcolor{darkgreen}{{\cal Y}_i}, \color{darkred}{\cal X}$ to be boundary-less. Here instead of ${\cal B}, {\cal C}$ we have used $\color{darkred}{\cal X}$ ($\color{darkgreen}{\cal Y}_i$) to convey the point that they can be timelike (spacelike) too, while in the figure we have depicted spacelike (timelike) case. $\color{darkpink}{\cal X}_{12}$ is the part of $\color{darkred}{\cal X}$ bounded between $\color{darkgreen}{\cal Y}_i$ and they  intersect on codimension-2 surfaces $\color{Blue}{\cal S}_i$. }
    \label{fig:3}
\end{figure}

\bec
\textbf{\tcm{Proof of proposition II.}}
\eec
Let us start with the Noether charge variation. To do so, we take variation over the last line of \eqref{proof-I}
\begin{equation}\label{proof-II}
   \begin{split}
        &\delta Q_{\text{\tiny{N}}}(\xi;\mathcal{S}_2)-\delta Q_{\text{\tiny{N}}}(\xi;\mathcal{S}_1)  =\delta \int_{\mathcal{X}_{12}}\d{}x_{\mu}\, \Theta^{\mu}[\delta_{\xi}\Phi;\Phi]\\
        &=-\delta \int_{\mathcal{M}}\d{}^{D}x\, \partial_{\mu}\mathcal{X}\, (\Theta_{\text{\tiny{LW}}}^{\mu}[\delta_{\xi}\Phi;\Phi]-\xi^{\mu}\, \mathrm{L})\, H(\mathcal{Y}_1)\, H(-\mathcal{Y}_2)\, \Delta (\mathcal{X})\\
        &=\int_{\mathcal{M}}\d{}^{D}x\, \Big[-\partial_{\mu}\mathcal{X}\, (\delta\Theta_{\text{\tiny{LW}}}^{\mu}[\delta_{\xi}\Phi;\Phi]{-\delta(\xi^{\mu}\, \mathrm{L})})\, H(\mathcal{Y}_1)\, H(-\mathcal{Y}_2)\, \Delta (\mathcal{X})\\
        &+ \Theta_{\text{\tiny{LW}}}^{\mu}[\delta_{\xi}\Phi;\Phi]\, \delta \left(\partial_{\mu}\mathcal{X}\, H(\mathcal{Y}_1)\, H(-\mathcal{Y}_2)\, \Delta (\mathcal{X})\right)\\
        &+\xi^{\mu}\, \Big(\partial_{\mu}\delta\mathcal{X}\, H(\mathcal{Y}_1)\, H(-\mathcal{Y}_2)\, \Delta (\mathcal{X})+\partial_{\mu}\mathcal{X}\, \delta \mathcal{Y}_1\, \Delta(\mathcal{Y}_1)\, H(-\mathcal{Y}_2)\, \Delta (\mathcal{X})\\
        &-\partial_{\mu}\mathcal{X}\, \delta \mathcal{Y}_2 H(\mathcal{Y}_1)\, \Delta(\mathcal{Y}_2)\, \Delta (\mathcal{X})+\partial_{\mu}\mathcal{X}\, \delta \mathcal{X}\, H(\mathcal{Y}_1)\, H(-\mathcal{Y}_2)\, \Delta' (\mathcal{X})\Big)\mathrm{L}\Big]\, ,
   \end{split}
\end{equation}
where if $\mathcal{X}_{12}$ to be a timelike/spacelike hypersurface then $\mathcal{Y}_i$ with $i=1,2$ are two spacelike/timelike hypersurfaces. From now on we will assume $\xi$ is a Killing vector. Then we will have $\Theta_{\text{\tiny{LW}}}^{\mu}[\delta_{\zeta}\Phi;\Phi]=0$ and 
\begin{equation}
\begin{split}
    \delta\Theta_{\text{\tiny{LW}}}^{\mu}[\delta_{\zeta}\Phi;\Phi]&=\Theta_{\text{\tiny{LW}}}^{\mu}[\delta_{\zeta}\delta\Phi;\Phi]{+\Theta_{\text{\tiny{LW}}}^{\mu}[\delta_{\delta\zeta}\Phi;\Phi]}\\
    &=\mathcal{L}_{\zeta} \Theta_{\text{\tiny{LW}}}^{\mu}[\delta\Phi;\Phi]{+\Theta_{\text{\tiny{LW}}}^{\mu}[\delta_{\delta\zeta}\Phi;\Phi]}\\
    &=\partial_{\nu}(\zeta^{\nu}\, \Theta_{\text{\tiny{LW}}}^{\mu})-\Theta_{\text{\tiny{LW}}}^{\nu}\partial_{\nu}\zeta^{\mu}{+\Theta_{\text{\tiny{LW}}}^{\mu}[\delta_{\delta\zeta}\Phi;\Phi]}\, .
\end{split}
\end{equation}
Using the above equation and $\delta \mathrm{L}=\partial_{\mu}\Theta_{\text{\tiny{LW}}}^{\mu}[\delta\Phi;\Phi]$, we find
\begin{align}
        \delta\Theta_{\text{\tiny{LW}}}^{\mu}[\delta_{\zeta}\Phi;\Phi]{-\delta(\zeta^{\mu}\, \mathrm{L})} & =2\partial_{\nu}(\zeta^{[\nu}\, \Theta_{\text{\tiny{LW}}}^{\mu]}){+\Theta_{\text{\tiny{LW}}}^{\mu}[\delta_{\delta\zeta}\Phi;\Phi]-\delta\zeta^{\mu}\, \mathrm{L}}\, \nonumber\\
         &=2\partial_{\nu}(\zeta^{[\nu}\, \Theta_{\text{\tiny{LW}}}^{\mu]})+J^{\mu}_{\delta \zeta}\, \\
         &=2\partial_{\nu}(\zeta^{[\nu}\, \Theta_{\text{\tiny{LW}}}^{\mu]})+\partial_{\nu}Q_{\text{\tiny{N}}}^{\mu\nu}(\delta \zeta)\nonumber
\end{align}

Substituting these results in \eqref{proof-II}, we reach
\begin{equation}
    \begin{split}
        &\delta Q_{\text{\tiny{N}}}(\zeta;\mathcal{S}_2)-\delta Q_{\text{\tiny{N}}}(\zeta;\mathcal{S}_1)\\
        &=\int_{\mathcal{M}}\d{}^{D}x\, \Big[{-\partial_{\mu}\mathcal{X}\, \partial_{\nu}Q_{\text{\tiny{N}}}^{\mu\nu}(\delta \zeta)\, H(\mathcal{Y}_1)\, H(-\mathcal{Y}_2)\, \Delta (\mathcal{X})} \\
        &\hspace{1.5 cm}+2\partial_{\mu}\mathcal{X}\,\partial_{\nu}(\zeta^{[\mu}\, \Theta_{\text{\tiny{LW}}}^{\nu]})\, H(\mathcal{Y}_1)\, H(-\mathcal{Y}_2)\, \Delta (\mathcal{X})\\
        &\hspace{1.5 cm}+\zeta^{\mu}\, \mathrm{L}\Big(\partial_{\mu}\mathcal{X}\, \delta \mathcal{Y}_1\, \Delta(\mathcal{Y}_1)\, H(-\mathcal{Y}_2)\, \Delta (\mathcal{X})\\
        &\hspace{1.5 cm}-\partial_{\mu}\mathcal{X}\, \delta \mathcal{Y}_2 H(\mathcal{Y}_1)\, \Delta(\mathcal{Y}_2)\, \Delta (\mathcal{X})\\
        &\hspace{1.5 cm}+\partial_{\mu}(\delta\mathcal{X}\, \Delta (\mathcal{X}))\, H(\mathcal{Y}_1)\, H(-\mathcal{Y}_2)\Big)\Big]\, .
    \end{split}
\end{equation}
Integrating by parts on the first, second, and fourth lines and noting that $\partial_{\mu}(\zeta^{\mu}\, \mathrm{L})=\mathcal{L}_{\zeta}\mathrm{L}=0$, we obtain
\begin{equation}
    \begin{split}
       \hspace{-5mm}\delta Q_{\text{\tiny{N}}}(\zeta;\mathcal{S}_2)&-\delta Q_{\text{\tiny{N}}}(\zeta;\mathcal{S}_1)         \\ 
        \hspace{-4mm}=\int_{\mathcal{M}}\d{}^{D}x\, \Big[&+\partial_{\mu}\mathcal{X}\, Q_{\text{\tiny{N}}}^{\mu\nu}(\delta \zeta)\, \partial_{\nu}\mathcal{Y}_1 \, \Delta(\mathcal{Y}_1)\, H(-\mathcal{Y}_2)\, \Delta (\mathcal{X})\\
        &-\partial_{\mu}\mathcal{X}\, Q_{\text{\tiny{N}}}^{\mu\nu}(\delta \zeta)\, \partial_{\nu}\mathcal{Y}_2 \, H(\mathcal{Y}_1)\, \Delta(-\mathcal{Y}_2)\, \Delta (\mathcal{X}) \\
        &-2\partial_{\mu}\mathcal{X}\,(\zeta^{[\mu}\, \Theta_{\text{\tiny{LW}}}^{\nu]})\, \partial_{\nu}\mathcal{Y}_1\, \Delta(\mathcal{Y}_1)\, H(-\mathcal{Y}_2)\, \Delta (\mathcal{X})\\
        &+2\partial_{\mu}\mathcal{X}\,(\zeta^{[\mu}\, \Theta_{\text{\tiny{LW}}}^{\nu]})\, \partial_{\nu}\mathcal{Y}_2\, H(\mathcal{Y}_1)\, \Delta(\mathcal{Y}_2)\, \Delta (\mathcal{X})\\
        &+\zeta^{\mu}\, \mathrm{L}\Big(\partial_{\mu}\mathcal{X}\, \delta \mathcal{Y}_1\, \Delta(\mathcal{Y}_1)\, H(-\mathcal{Y}_2)\, \Delta (\mathcal{X})\\
        &-\partial_{\mu}\mathcal{X}\, \delta \mathcal{Y}_2 H(\mathcal{Y}_1)\, \Delta(\mathcal{Y}_2)\, \Delta (\mathcal{X})\\
        &-\delta\mathcal{X}\, \partial_{\mu}\mathcal{Y}_1 \Delta (\mathcal{X})\, \Delta(\mathcal{Y}_1)\, H(-\mathcal{Y}_2)\\
        &+\delta\mathcal{X}\, \partial_{\mu}\mathcal{Y}_2 \Delta (\mathcal{X})\, H(\mathcal{Y}_1)\, \Delta(\mathcal{Y}_2)\Big)\Big]\, .
    \end{split}
\end{equation}
All terms in the above equation are corner terms and we find
\begin{equation}
    \begin{split}
        &\delta Q_{\text{\tiny{N}}}(\zeta;\mathcal{S}_2)-\delta Q_{\text{\tiny{N}}}(\zeta;\mathcal{S}_1)        \\& \hspace{-4mm} =\int_{\mathcal{S}_1}\d{}^{D-2}x\, \Big[{\partial_{[\mu}\mathcal{X}\, \partial_{\nu]}\mathcal{Y}_1 \, Q_{\text{\tiny{N}}}^{\mu\nu}(\delta \zeta)}+2\partial_{[\nu}\mathcal{X}\, \partial_{\mu]}\mathcal{Y}_1\, \zeta^{\mu}\, \Theta_{\text{\tiny{LW}}}^{\nu}\\
        &\hspace{15mm}+\zeta^{\mu}\, \mathrm{L}\big(\partial_{\mu}\mathcal{X}\, \delta \mathcal{Y}_1 -\delta\mathcal{X}\, \partial_{\mu}\mathcal{Y}_1\big)\Big]\\
        &\hspace{-4mm}-\int_{\mathcal{S}_2}\d{}^{D-2}x\, \Big[{\partial_{[\mu}\mathcal{X}\, \partial_{\nu]}\mathcal{Y}_2 \, Q_{\text{\tiny{N}}}^{\mu\nu}(\delta \zeta)}+2\partial_{[\nu}\mathcal{X}\, \partial_{\mu]}\mathcal{Y}_2\, \zeta^{\mu}\, \Theta_{\text{\tiny{LW}}}^{\nu}\\
        &\hspace{15mm}+\zeta^{\mu}\, \mathrm{L}\big(\partial_{\mu}\mathcal{X}\, \delta \mathcal{Y}_2-\delta\mathcal{X}\, \partial_{\mu}\mathcal{Y}_2\big)\Big]\, .
    \end{split}
\end{equation}
Finally,  using \eqref{chi-chi*-def} and \eqref{notation-int}, we arrive at
\begin{equation}
        \begin{split}
        \delta Q_{\text{\tiny{N}}}(\zeta;\mathcal{S}_2)&-\delta Q_{\text{\tiny{N}}}(\zeta;\mathcal{S}_1)={Q_{\text{\tiny{N}}}(\delta\zeta;\mathcal{S}_2)-Q_{\text{\tiny{N}}}(\delta\zeta;\mathcal{S}_2)}\\
        &-2\int_{\mathcal{S}_2}\d{}x_{\mu\nu}\, \zeta^{\mu}\, \Theta^{\nu}
        +2\int_{\mathcal{S}_1}\d{}x_{\mu\nu}\, \zeta^{\mu}\, \Theta^{\nu}\, .
        \end{split}
\end{equation}
The above  yields $K(\zeta; \mathcal{S}_1)=K(\zeta; \mathcal{S}_2)$ with $K$ is given by \eqref{K-delta-nonzero}.

\bec
\textbf{\tcm{Proof of Proposition III.}}
\eec
We start from \eqref{Noether-cons} and rewrite it for the $W,Y$ shifted Noether charge $\tilde{Q}_{\text{\tiny{N}}}$ \eqref{charge-freedom} and Lagrangian $\tilde{\mathrm{L}}[\Phi]$:
\begin{equation}\label{freedom-Q-check}
\begin{split}
    \tilde{Q}_{\text{\tiny{N}}}(\zeta;\mathcal{S}_{2})-\tilde{Q}_{\text{\tiny{N}}}(\zeta;\mathcal{S}_1)&=-\int_{\mathcal{X}_{12}}\d{}x_{\mu}\, \zeta^{\mu}\, \tilde{\mathrm{L}}[\Phi]\, ,
    \end{split}
\end{equation}
where $\tilde{\mathrm{L}}[\Phi]=\mathrm{L}[\Phi]+\partial_\mu W^\mu$ and 
\begin{equation}\begin{split}
    \tilde{Q}_{\text{\tiny{N}}}(\zeta)&=Q_{\text{\tiny{N}}}(\zeta)+\int_{\cal S}\d{}x_{\mu\nu}\, \bar{Y}^{\mu\nu}[\delta_{\zeta}\Phi;\Phi]\\
    &=Q_{\text{\tiny{N}}}(\zeta)+2\int_{\cal S}\d{}x_{\mu\nu}\, W^{\mu}\, \zeta^{\nu}\, ,
\end{split}\end{equation}
where we have used 
\begin{equation}\label{Y-bar-zeta}
    \bar{Y}^{\mu\nu}[\delta_{\zeta}\Phi;\Phi]={Y}^{\mu\nu}[\delta_{\zeta}\Phi;\Phi]+2 W^{[\mu}\, \zeta^{\nu]}=2 W^{[\mu}\, \zeta^{\nu]}\, ,
\end{equation}
and $Y^{\mu\nu}[\delta_{\zeta}\Phi;\Phi]=0$ for exact symmetry $\zeta$.
The LHS of  \eqref{freedom-Q-check} can be written as follows
\begin{align}\label{Freedom-QN-cons-1}
    &\hspace{-3mm}\tilde{Q}_{\text{\tiny{N}}}(\zeta;\mathcal{S}_{2})-\tilde{Q}_{\text{\tiny{N}}}(\zeta;\mathcal{S}_1)\nonumber \\
    &\hspace{-3mm}=Q_{\text{\tiny{N}}}(\zeta;\mathcal{S}_{2})-Q_{\text{\tiny{N}}}(\zeta;\mathcal{S}_{1})+2\int_{{\cal S}_2}\d{}x_{\mu\nu}\, W^{\mu}\, \zeta^{\nu}-2\int_{{\cal S}_1}\d{}x_{\mu\nu}\, W^{\mu}\, \zeta^{\nu}\, \nonumber \\
    &\hspace{-3mm}=Q_{\text{\tiny{N}}}(\zeta;\mathcal{S}_{2})-Q_{\text{\tiny{N}}}(\zeta;\mathcal{S}_{1})+2\int_{\mathcal{X}_{12}}\d{}x_{[\mu}\, \partial_{\nu]} \big(W^{\mu}\, \zeta^{\nu}\big)\,\\
    &\hspace{-3mm}=Q_{\text{\tiny{N}}}(\zeta;\mathcal{S}_{2})-Q_{\text{\tiny{N}}}(\zeta;\mathcal{S}_{1})+2\int_{\mathcal{X}_{12}}\d{}x_{\mu}\, \partial_{\nu} \big(W^{[\mu}\, \zeta^{\nu]}\big)\, .\nonumber 
\end{align}
Let us now explore the right-hand side of the \eqref{freedom-Q-check},  using
\begin{equation}
    \int_{\mathcal{X}_{12}} \d{x}_{\mu}\, \zeta^{\mu}\, \tilde{\mathrm{L}}=\int_{\mathcal{X}_{12}} \d{x}_{\mu}\, \zeta^{\mu}\, {\mathrm{L}}+\int_{\mathcal{X}_{12}} \d{x}_{\mu}\, \zeta^{\mu}\, \partial_{\nu}W^{\nu}\, ,
\end{equation}
and that $\mathcal{L}_{\zeta}W=0$, we deduce $\zeta^{\mu}\, \partial_\nu W^\nu=- 2 \partial_{\nu}(W^{[\mu}\, \zeta^{\nu]})$. Then 
\begin{equation}\label{Freedom-QN-cons-2}
    \int_{\mathcal{X}_{12}} \d{x}_{\mu}\, \zeta^{\mu}\, \tilde{\mathrm{L}}=\int_{\mathcal{X}_{12}} \d{x}_{\mu}\, \zeta^{\mu}\, {\mathrm{L}}-2\int_{\mathcal{X}_{12}} \d{x}_{\mu}\, \partial_{\nu}(W^{[\mu}\, \zeta^{\nu]})\, .
\end{equation}
Putting together LHS \eqref{Freedom-QN-cons-1} and RHS \eqref{Freedom-QN-cons-2}, we get
\begin{equation}
\begin{split}
    {Q}_{\text{\tiny{N}}}(\zeta;\mathcal{S}_{2})-{Q}_{\text{\tiny{N}}}(\zeta;\mathcal{S}_1)&=-\int_{\mathcal{X}_{12}}\d{}x_{\mu}\, \zeta^{\mu}\, \mathrm{L}[\Phi]\, ,
    \end{split}
\end{equation}
which shows that the Noether conservation law \eqref{Noether-cons} is independent of $W$ and $Y$ freedoms/ambiguities.

\bec
\textbf{\tcm{Proof of Proposition IV.}}
\eec
Let us start from the $W, Y$ transformed  $K(\xi;\cal S)$, denoted as  $\tilde{K}(\xi;\mathcal{S})$,
    \begin{align}
     \hspace{-7mm}   \tilde{K}(\xi;\mathcal{S}) =&-\delta \tilde{Q}_{\text{\tiny{N}}}(\xi){+\tilde{Q}_{\text{\tiny{N}}}(\delta\xi)}+2\int_{\mathcal{S}}\d{}x_{\mu\nu}\, \xi^{\mu}\, \tilde{\Theta}^{\nu}\nonumber \\
         =& K(\xi;\mathcal{S})+2\int_{\mathcal{S}}\d{}x_{\mu\nu}\, \xi^{\mu}\, ( \delta W^\nu + \chi^\nu\, \partial_\alpha W^\alpha+\partial_{\alpha}Y^{\nu\alpha})\nonumber\\
        &+\delta \int_{{\cal S}}\d{}x_{\mu\nu}\, \bar{Y}^{\mu\nu}[\delta_{\xi}\Phi;\Phi]
         {-\int_{\cal S}\d{}x_{\mu\nu}\, \bar{Y}^{\mu\nu}[\delta_{\delta\xi}\Phi;\Phi]}\\
         =& {K}(\xi;\mathcal{S})+ \int_{{\cal S}} \d{}x_{\mu\nu}\, \Big[2\, \xi^{\mu}\, ( \delta W^\nu + \chi^\nu\, \partial_\alpha W^\alpha+\partial_{\alpha}Y^{\nu\alpha}) \nonumber\\
         &+\delta\bar{Y}^{\mu\nu}[\delta_{\xi}\Phi;\Phi]{-\bar{Y}^{\mu\nu}[\delta_{\delta\xi}\Phi;\Phi]}+2\, \chi^{\nu}\, \partial_{\alpha}\bar{Y}^{\mu\alpha}[\delta_{\xi}\Phi;\Phi]\Big]\nonumber
    \end{align}
In the second equality we have used \eqref{symp-pot-freedom} and \eqref{charge-freedom} and in the third, the following identity \cite{Adami:2024gdx}
\begin{equation}
    \delta \int_{\mathcal{S}} \d{}x_{\mu\nu}\, X^{\mu\nu}=\int_{\mathcal{S}} \d{}x_{\mu\nu}\, (\delta X^{\mu\nu}+2\,\chi^{\nu}\, \partial_{\alpha}X^{\mu\alpha})\, ,
\end{equation}
where $X^{\mu\nu}$ is a {bi-vector} density in spacetime and a 0-form in phase space.
Recalling \eqref{Y-bar-zeta} and 
\begin{equation}\begin{split}
        & \hspace{-5mm}\delta\bar{Y}^{\mu\nu}[\delta_{\zeta}\Phi;\Phi] {-\bar{Y}^{\mu\nu}[\delta_{\delta\zeta}\Phi;\Phi]}
        \\ &={Y}^{\mu\nu}[\delta_{\zeta}\delta\Phi;\Phi]+2 \delta W^{[\mu}\, \zeta^{\nu]}\,
        \\&=\mathcal{L}_{\zeta}{Y}^{\mu\nu}[\delta\Phi;\Phi]+2 \delta W^{[\mu}\, \zeta^{\nu]}\,\\
        & =\partial_{\alpha}(\zeta^{\alpha}\, {Y}^{\mu\nu})-{Y}^{\alpha\nu}\, \partial_{\alpha} \zeta^{\mu}-{Y}^{\mu\alpha}\, \partial_{\alpha} \zeta^{\nu}+2 \delta W^{[\mu}\, \zeta^{\nu]}\, ,
\end{split}\end{equation}
{where we used
        $\bar{Y}^{\mu\nu}[\delta_{\delta\zeta}\Phi;\Phi]={Y}^{\mu\nu}[\delta_{\delta\zeta}\Phi;\Phi]+2 W^{[\mu}\, \delta\zeta^{\nu]}$,}
we learn 
\begin{equation}
   \begin{split}
    \tilde{K}&(\zeta;\mathcal{S}) =K(\zeta;\mathcal{S})
    \\ &+ \int_{{\cal S}}\d{}x_{\mu\nu}\; \Big[4\, \chi^{\nu}\, \partial_{\alpha}(W^{[\mu}\, \zeta^{\alpha]})+2\, \zeta^{\mu}\, \chi^\nu\, \partial_\alpha W^\alpha \\ &+  \partial_{\alpha}(\zeta^{\alpha}\, {Y}^{\mu\nu})-{Y}^{\alpha\nu}\, \partial_{\alpha} \zeta^{\mu}-{Y}^{\mu\alpha}\, \partial_{\alpha} \zeta^{\nu}+2\, \zeta^{\mu}\, \partial_{\alpha}Y^{\nu\alpha}\Big]\, .
   \end{split}
\end{equation}
The first and the second lines of the integrand vanish. To see this we note that 
\begin{equation*}\begin{split}
    \text{First Line}&=2 \partial_{\alpha}(W^{[\mu}\, \zeta^{\alpha]})+ \zeta^{\mu}\, \partial_\alpha W^\alpha\\&= \partial_\alpha(W^\mu \zeta^{\alpha})-W^\alpha\partial_\alpha \zeta^{\mu} = \mathcal{L}_{\zeta} W^\mu=0\, ,
\end{split}\end{equation*}
and
\begin{equation*}\begin{split}
    \text{Second Line}&=\int_{{\cal S}}\d{}x_{\mu\nu}\,\Big[\partial_{\alpha}(\zeta^{\alpha}\, {Y}^{\mu\nu})-{Y}^{\alpha\nu}\, \partial_{\alpha} \zeta^{\mu}\\ 
    &\hspace{16mm}-{Y}^{\mu\alpha}\, \partial_{\alpha} \zeta^{\nu}+2\, \zeta^{\mu}\, \partial_{\alpha}Y^{\nu\alpha}\Big]\\
    &=\int_{{\cal S}}\d{}x_{\mu\nu}\,\Big[\partial_{\alpha}(\zeta^{\alpha}\, {Y}^{\mu\nu})+2{Y}^{\nu\alpha}\, \partial_{\alpha} \zeta^{\mu}+2\, \zeta^{\mu}\, \partial_{\alpha}Y^{\nu\alpha}\Big]\\
    &=\int_{{\cal S}}\d{}x_{\mu\nu}\,\Big[\partial_{\alpha}(\zeta^{\alpha}\, {Y}^{\mu\nu})+2 \partial_{\alpha}({Y}^{\nu\alpha}\, \zeta^{\mu})\Big]\\
    &=3\int_{{\cal S}}\d{}x_{\mu\nu}\,\partial_{\alpha}\big(\zeta^{[\alpha}\, {Y}^{\mu\nu]}\big)\\
    &=\frac{3}{2}\int_{{\cal S}}\d{}^{D-2}x\, \partial_{\alpha}\Big(\mathcal{E}_{\mu\nu}\,\zeta^{[\alpha}\, {Y}^{\mu\nu]}\Big)=0 \, .
\end{split}\end{equation*}
In the last line, we used $\partial_{[\alpha}{\cal E}_{\mu\nu]}=0$. 
Finally, we arrive at the desired result 
\begin{equation}
    \tilde{K}(\zeta;\mathcal{S})  =K(\zeta;\mathcal{S})\, .
\end{equation}
\bec
\textbf{Another Form of Smarr Formula}\label{app:smarr-2}
\eec
Our derivation of generalized Smarr formula \eqref{temp-rad-cons} involves Noether charges and is invariant under $W$-freedom. One may hence add an arbitrary $W$ term to both sides of our Smarr relations. One such $W$ choice is the one that relates Noether mass $M_{\text{\tiny{N}}}$ to the ADM mass $M$ \cite{Iyer:1994ys, Iyer:1995kg, Wald:1999wa}: 
\begin{equation}
   K(\zeta_t;\mathcal{S}_{i^0})=\delta \int_{\mathcal{S}}\d{}x_{\mu\nu}\, Q^{\mu\nu}_{\text{\tiny{N}}}(\zeta_t)+2\int_{\mathcal{S}}\d{}x_{\mu\nu}\, \zeta_t^{\mu}\, \Theta^{\nu}
\end{equation}
assuming that there exists  a 0-form in the phase space $B^\mu$ such that
\begin{equation}\label{}
    \int_{\mathcal{S}_{i^0}}\d{}x_{\mu\nu}\, \zeta_t^{\mu}\, \Theta^{\nu}=\delta\, \int_{\mathcal{S}_{i^0}}\d{}x_{\mu\nu}\, \zeta_t^{\mu}\, B^{\nu}
\end{equation}
Then, we will have
\begin{equation}
    \delta M=\delta  Q_{\text{\tiny{N}}}(\zeta_t,\mathcal{S}_{i^0})+2\,\delta\,\int_{\mathcal{S}_{i^0}}\d{}x_{\mu\nu}\, \zeta_t^{\mu}\, B^{\nu}\, ,
\end{equation}
where $M$ is  the ADM mass (up to a reference point mass)
\begin{equation}
    M=  Q_{\text{\tiny{N}}}(\zeta_t,\mathcal{S}_{i^0})+2 \int_{\mathcal{S}_{i^0}}\d{}x_{\mu\nu}\, \zeta_t^{\mu}\, B^{\nu}\, ,
\end{equation}
Thus, the Smarr formula can be rewritten as follows
\begin{equation}
    -M+\Omega_{\text{\tiny{H}}}\, J+T_{\text{\tiny{H}}}\, S_{\text{\tiny{H}}}= T_{\text{\tiny{H}}}\int_{\cauchy_{12}}\d{}x_{\mu}\, \zeta_{\text{\tiny{H}}}^{\mu}\, \mathrm{L}[\Phi]-2 \int_{\mathcal{S}_{i^0}}\d{}x_{\mu\nu}\, \zeta_t^{\mu}\, B^{\nu}\, . 
\end{equation}
We can write the last term on the RHS as follows
\begin{align}
        2 \int_{\mathcal{S}_{i^0}}\d{}x_{\mu\nu}\, \zeta_t^{\mu}\, B^{\nu}&=2 T_{\text{\tiny{H}}}\int_{\mathcal{S}_{i^0}}\d{}x_{\mu\nu}\, \zeta_{\text{\tiny{H}}}^{\mu}\, B^{\nu}-2T_{\text{\tiny{H}}} \int_{\mathcal{S}_{_{b}}}\d{}x_{\mu\nu}\, \zeta_{\text{\tiny{H}}}^{\mu}\, B^{\nu}\nonumber \\
        &=2T_{\text{\tiny{H}}}\, \int_{\mathcal{C}_{12}}\d{}x_{[\mu}\partial_{\nu]}(\zeta_{\text{\tiny{H}}}^{\mu}\, B^{\nu})\\
        &=2T_{\text{\tiny{H}}}\, \int_{\mathcal{C}_{12}}\d{}x_{\mu}\partial_{\nu}(\zeta_{\text{\tiny{H}}}^{[\mu}\, B^{\nu]})\nonumber 
\end{align}
Since $\mathcal{L}_{\zeta}B^{\mu}=0$, we deduce $\zeta^{\mu}\, \partial_\nu B^\nu=- 2 \partial_{\nu}(B^{[\mu}\, \zeta^{\nu]})$ and 
\begin{equation}
    2 \int_{\mathcal{S}_{i^0}}\d{}x_{\mu\nu}\, \zeta_t^{\mu}\, B^{\nu}=T_{\text{\tiny{H}}}\int_{\mathcal{C}_{12}}\d{}x_{\mu}\, \zeta_{\text{\tiny{H}}}^{\mu}\, \partial_\nu B^\nu\, ,
\end{equation}
yielding the Smarr formula in terms of ADM mass $M$ \eqref{Smarr-3}
\begin{equation}\label{Smarr-2}
           \hspace{-.35 cm} -M+\Omega_{\text{\tiny{H}}}\, J+T_{\text{\tiny{H}}}\, S_{\text{\tiny{H}}}= T_{\text{\tiny{H}}}\int_{\cauchy_{12}}\d{}x_{\mu}\, \zeta_{\text{\tiny{H}}}^{\mu}\, (\mathrm{L}[\Phi]- \partial_\nu B^\nu)\,. \hspace{-.35 cm} 
\end{equation}


\bec
\textbf{Smarr Relation for (A)dS-Myers-Perry Black Holes}
\eec
For the $D$-dimensional case with $D=2n+1+\alpha$ (with $\alpha= 0, 1$ respectively for odd and even $D$) and $\Lambda=-(D-1)(D-2)/(2L^2)$ we have the solution \cite{Gibbons:2004uw, Gibbons:2004js}
\begin{align}
        \d{}s^2&= -W(1+L^{-2}\, r^{2})\d{}t^2+\frac{2m}{V\, F}\left(W\, \d{}t-\sum_{i=1}^{n}\frac{a_i}{\Xi_i}\mu_{i}^2\, \d{}\phi_{i}\right)^2\nonumber \\
        &+\sum_{i=1}^{n}\frac{r^2+a_i^2}{\Xi_i}\mu_{i}^2\, \d{}\phi_i^2+\frac{V\, F}{V-2m}\d{}r^2+\sum_{i=1}^{n+\alpha}\frac{r^2+a_i^2}{\Xi_i}\d{}\mu_i^2\nonumber\\
        &-\frac{1}{W(L^2+r^2)}\left(\sum_{i=1}^{n+\alpha}\frac{r^2+a_i^2}{\Xi_i}\mu_i\, \d{}\mu_i\right)^2\, ,
\end{align}
where
\begin{equation}
   \begin{split}
        &\Xi_i=1-L^{-2}\, a_{i}^2\, , \quad W=\sum_{i=1}^{n+\alpha}\frac{\mu_i^2}{\Xi_i}\, , \quad \sum_{i=1}^{n+\alpha}\mu_i^2=1\, ,\\
        &V=r^{\alpha-2}(1+L^{-2}\,r^2)\prod_{i=1}^{n}(r^2+a_i^2)\, ,\\
        &F=\frac{r^2}{1+L^{-2}\,r^2}\sum_{i=1}^{n+\alpha}\frac{\mu_i^2}{r^2+a_i^2}\, .
   \end{split}
\end{equation}
The Killing horizon generator is
\begin{equation}
\zeta_{\text{\tiny{H}}}=\frac{1}{T_{\text{\tiny{H}}}}\left(\partial_{t}+\sum_{i=1}^{n}\Omega^i_{\text{\tiny{H}}}\, \partial_{\phi^i}\right), \qquad 
   \Omega^i_{\text{\tiny{H}}}= \frac{a_i(1+L^{-2}\, r_{\text{\tiny{H}}}^2)}{r_{\text{\tiny{H}}}^2+a_i^2}\, ,
\end{equation}
where $r_{\text{\tiny{H}}}$ is the simple root of $V(r)-2m=0$ and the black hole (Hawking) temperature is given by
\begin{equation}
    T_{\text{\tiny{H}}}=\frac{1}{2\pi}\left[r_{\text{\tiny{H}}}(1+L^{-2}\,r_{\text{\tiny{H}}}^2)\left(\sum_{i=1}^{n}\frac{1}{r_{\text{\tiny{H}}}^2+a_i^2}+\frac{\alpha}{2r_{\text{\tiny{H}}}^2}\right)-\frac{1}{r_{\text{\tiny{H}}}}\right]\, .
\end{equation}
The ADM mass $K(\zeta_t;\mathcal{S}_{i^0})=\delta M$ and angular momentum $Q_{\text{\tiny{N}}}(\zeta_{\phi_i};\mathcal{S}_{i^0})=-J_i$ are given by
\begin{equation}
\begin{split}
\hspace*{-4mm} M=\frac{m\, \Omega_{D-2}}{4\pi\, G \prod_{_j}\Xi_j}\left(\sum_{i=1}^n\frac{1}{\Xi_i}+\frac{\alpha-1}{2}\right), \,
    J_i=\frac{m\, \Omega_{D-2}}{4\pi\, G \prod_{_j}\Xi_j}\frac{a_i}{\Xi_i}.
\end{split}
\end{equation}
{This result was found in \cite{Gibbons_2005}.}
To explore the Smarr formula consider the Noether mass
\begin{equation}
     \begin{split}
         M_{\text{\tiny{N}}}&=M\left(1 -\frac{1}{2}\frac{1}{\sum_{i=1}^n\frac{1}{\Xi_i}+\frac{\alpha-1}{2}} \right)\\
         &-\frac{\Lambda\, \Omega_{D-2}}{4\pi\,G}\frac{r_{\infty}^{\alpha}\prod_{i=1}^{n}(r_{\infty}^2+a_i^2)}{(D-1)(D-2)\prod_{_j}\Xi_j}\, ,
     \end{split}
\end{equation}
and
\begin{equation}
     \begin{split}
         T_{\text{\tiny{H}}}\int_{\cauchy_{12}}\d{}x_{\mu}\, \zeta_{\text{\tiny{H}}}^{\mu}\, \mathrm{L}[\Phi]&=-\frac{\Lambda\, \Omega_{D-2}}{4\pi\,G}\frac{r_{\text{\tiny{H}}}^{\alpha}\prod_{i=1}^{n}(r_{\text{\tiny{H}}}^2+a_i^2)}{(D-1)(D-2)\prod_{_j}\Xi_j}\\
         &+\frac{\Lambda\, \Omega_{D-2}}{4\pi\,G}\frac{r_{\infty}^{\alpha}\prod_{i=1}^{n}(r_{\infty}^2+a_i^2)}{(D-1)(D-2)\prod_{_j}\Xi_j}\, .
     \end{split}
\end{equation}
Finally, the Smarr formula \eqref{Noether-cons} is given by 
\begin{equation}
    \begin{split}
        &-M\left(1 -\frac{1}{2}\frac{1}{\sum_{i=1}^n\frac{1}{\Xi_i}+\frac{\alpha-1}{2}} \right)+\sum_{i=1}^{n}\Omega^{i}_{\text{\tiny{H}}}\, J_i+T_{\text{\tiny{H}}}\, S_{\text{\tiny{H}}}\\
        &= -\frac{\Lambda\, \Omega_{D-2}}{4\pi\,G}\frac{r_{\text{\tiny{H}}}^{\alpha}\prod_{i=1}^{n}(r_{\text{\tiny{H}}}^2+a_i^2)}{(D-1)(D-2)\prod_{_j}\Xi_j}\, .
    \end{split}
\end{equation}
This result was also found in \cite{Barnich:2004uw}. 
In the flat limit, we get the standard result for the Myers-Perry black holes \cite{Myers:1986un, Myers:2011yc}
\begin{equation}
    -\frac{D-3}{D-2}M+\sum_{i=1}^{n}\Omega^{i}_{\text{\tiny{H}}}\, J_i+T_{\text{\tiny{H}}}\, S_{\text{\tiny{H}}}=0\, .
\end{equation}

\bibliography{reference}
\end{document}